\documentclass[reprint,aps,prl,superscriptaddress]{revtex4-2}
\usepackage{graphicx}
\usepackage{dcolumn}
\usepackage{bm}
\usepackage[usenames]{color}
\usepackage{xcolor}
\usepackage{tabularx}
\usepackage{booktabs}
\usepackage{hyperref}
\usepackage{siunitx}
\usepackage[utf8]{inputenc}
\usepackage[english]{babel}
\hypersetup{colorlinks,linkcolor=red,citecolor=blue,urlcolor=blue,final}
\usepackage{amsmath, amsthm, amscd, amssymb} 

\newcommand{\rvec}{{\bf r}}
\newcommand{\Fz}{F_z}
\newcommand{\Fr}{F_\rho}

\newcommand{\Tint}{T_i}

\begin{document}
\title{Levitated Optomechanics with Meta-Atoms}

\author{Sergei~Lepeshov}
\affiliation{School of Physics and Engineering, ITMO University, Saint Petersburg, Russia}

\author{Nadine~Meyer}
\affiliation{Nanophotonic Systems Laboratory, Department of Mechanical and Process Engineering, ETH Zurich, 8092 Zurich, Switzerland}
\affiliation{Quantum Center, ETH Zurich, 8083 Zurich, Switzerland}

\author{Patrick Maurer}
\affiliation{Institute for Quantum Optics and Quantum Information of the Austrian Academy of Sciences, A-6020 Innsbruck, Austria.}
\affiliation{Institute for Theoretical Physics, University of Innsbruck, A-6020 Innsbruck, Austria.}

\author{Oriol~Romero-Isart}
\affiliation{Institute for Quantum Optics and Quantum Information of the Austrian Academy of Sciences, A-6020 Innsbruck, Austria.}
\affiliation{Institute for Theoretical Physics, University of Innsbruck, A-6020 Innsbruck, Austria.}

\author{Romain~Quidant}
\affiliation{Nanophotonic Systems Laboratory, Department of Mechanical and Process Engineering, ETH Zurich, 8092 Zurich, Switzerland}
\affiliation{Quantum Center, ETH Zurich, 8083 Zurich, Switzerland}

\date{\today}
\begin{abstract} 

We propose to introduce additional control in levitated optomechanics by trapping a meta-atom, i.e. a subwavelength and high-permittivity dielectric particle supporting Mie resonances. In particular, we theoretically demonstrate that optical levitation and center-of-mass ground-state cooling of silicon nanoparticles in vacuum is not only experimentally feasible but it offers enhanced performance over widely used silica particles, in terms of trap frequency, trap depth and optomechanical coupling rates. Moreover, we show that, by adjusting the detuning of the trapping laser with respect to the particle’s resonance, the sign of the polarizability becomes negative, enabling levitation in the minimum of laser intensity e.g. at the nodes of a standing wave. The latter opens the door to trapping nanoparticles in the optical near-field combining red and blue-detuned frequencies, in analogy to two-level atoms, which is of interest for generating strong coupling to photonic nanostructures and short-distance force sensing.

\end{abstract}

\maketitle

Optical trapping and motional control of polarizable sub-micron objects in vacuum has become a very active research field~\cite{millen2020optomechanics, gonzalez2021levitodynamics}. Recently, the motion of an optically levitated silica nanoparticle has been cooled to the quantum ground state, either using passive feedback cooling via coherent scattering into a cavity~\cite{delic2020cooling,ranfagni2022two,piotrowski2022simultaneous}, or via active feedback cooling~\cite{tebbenjohanns2021quantum,magrini2021real,kamba2022optical} with shot-noise limited optical detection. In addition, the optical dipole-dipole interaction between two silica nanoparticles trapped in vacuum in two separate optical tweezers has been measured~\cite{rieser2022tunable}, which opens the door to study many-particle physics in vacuum~\cite{burnham2006holographic,dholakia2010colloquium,lechner2013cavity,liu2020prethermalization,gonzalez2021levitodynamics,yan2022ondemand, verkerk1992dynamics,jessen1992observation}. Optical trapping and control in vacuum of more complex particles supporting internal resonances has thus far been considered unattainable due to laser absorption, and have been solely studied by using low frequency electric~\cite{fonseca2016nonlinear,conangla2018motion,conangla2020extending,delord2020spin} or magnetic fields~\cite{wang2019dynamics,timberlake2019acceleration,vinante2020ultralow,delord2020spin,gieseler2020single,lewandowski2021high,latorre2022chip,latorre2022superconducting, romero2012quantum}. In contrast to optical manipulation, magnetic and electric platforms are less advanced and face additional challenges in terms of motional control in the quantum regime~\cite{gonzalez2021levitodynamics}. 

In this Letter, we analyze the use of Mie resonances supported by silicon nanoparticles~\cite{zhang2018lighting} for optical levitation in vacuum~\cite{ashkin1977observation}. We demonstrate that the resonances enable larger trap frequencies and trap depths compared to silica nanoparticles. Remarkably, we also evidence that silicon nanoparticles of few hundred nanometers behave, from the optomechanical standpoint, as a meta-atom whose polarizability changes sign across the resonance. In analogy with two-level atoms, this enables trapping the particle in regions of minimum laser intensity~\cite{grimm2000optical,kaplan2002optimized,jaouadi2010bose,Juan2016nearfield}.  The frequency-dependent sign of the polarizability is foreseen to enable trapping of silicon particles near a surface by using two-color near-field traps~\cite{LeKien2004atom,vetsch2010optical}. The latter is of interest to couple particle's motion to optical microcavities~\cite{vahala2003optical} or other integrated photonic systems~\cite{akahane2003high,deotare2009high,magrini2018near}. In the context of optical interaction of many particles, optical resonances pave the way to engineer stronger and more complex types of interactions beyond dipole-dipole interaction. Last but not least, in the context of levitated optomechanics, we theoretically show how to achieve motional ground-state cooling of optically resonant nanoparticles and discuss its distinctive features, including larger cooling rates and even the possibility to enter the strong optomechanical coupling regime in free space~\cite{ranfagni2021vectorial,sommer2021strong,delic2020cooling}.


\begin{figure}
\centering
\includegraphics[width=0.95\linewidth]{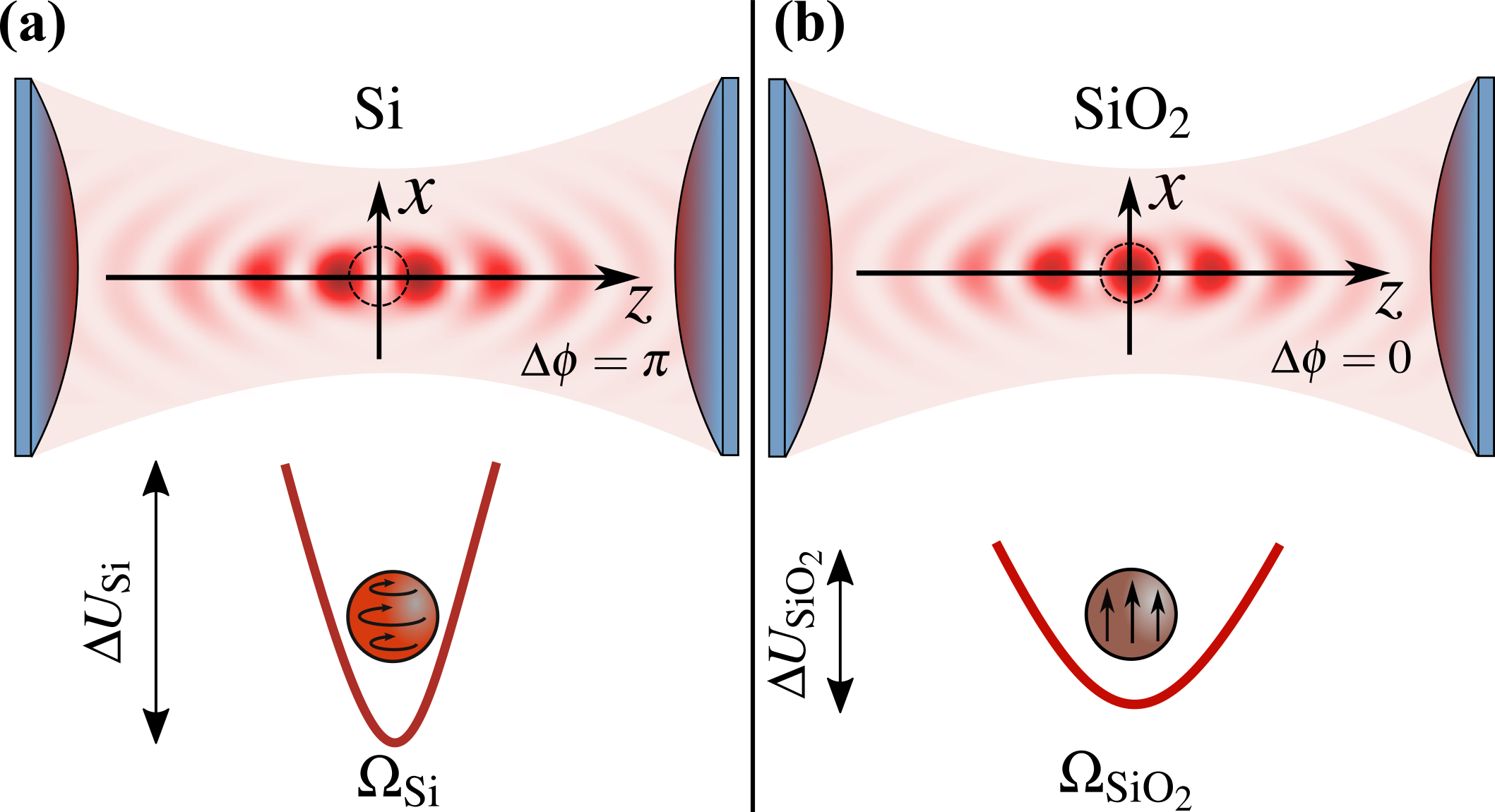}
\caption{Illustration of the optical configuration and the corresponding optical potential with trap depth $\Delta U$ and trap frequency $\Omega$. In both panels the optical configuration consists of two $x-$polarized focused laser beams of equal wavelength, counter-propagating along the $z-$axis. The relative phase $\Delta \phi$ specifies the intensity of the standing wave at the focal point. (a) Silicon nanosphere trapped at the intensity minimum ($\Delta \phi=\pi$) (b) Silica nanosphere trapped at the intensity maximum ($\Delta \phi = 0$).
}
 \label{fig:1}
\end{figure}

Let us consider a spherical silicon (Si)  particle of radius $R$, mass $m$, and homogeneous refractive index $n = n_r + \text{i} n_i$, interacting with laser light in ultra-high vacuum. We consider a standing wave pattern along the $z$-axis that is formed by two $x-$polarized and counter-propagating focused laser beams of equal wavelength $\lambda$, and with a relative phase $\Delta\phi$~\cite{vcivzmar2005optical,brzobohaty2013experimental,grass2016optical,nikkhou2021direct} (Fig.~\ref{fig:1}). The case $\Delta \phi =0$ and $\Delta \phi =\pi$ corresponds to the optical configuration with constructive interference and destructive interference at the focal point, respectively.  Due to the symmetry of the illumination, the scattering forces experienced by the particle cancel out such that trapping conditions are fully determined by both field intensity gradients and particle polarizability.  
Unlike subwavelength silica (SiO$_2$) nanoparticles that behave as non-resonant dipole scatterers, a subwavelength Si nanoparticle supports multipolar Mie resonances~\cite{zhang2018lighting} whose spectral features depend on the real part of the dielectric constant $\epsilon = n^2$ and the ratio $R/\lambda$. The enhanced nanoparticle polarizability at a given Mie resonance is expected to substantially increase the total force experienced by the particle, and thereby to increase trap depth and trap frequencies compared to a  SiO$_2$ nanoparticle (Fig.~\ref{fig:1}). Furthermore, as discussed later, Mie resonances offer an opportunity to control the sign of the force by an appropriate detuning between the Mie resonance and the frequency of the trapping laser, thereby enabling to trap particles at a dark spot (Fig.~\ref{fig:1}a). 


\begin{figure}
\centering
\includegraphics[width=\linewidth]{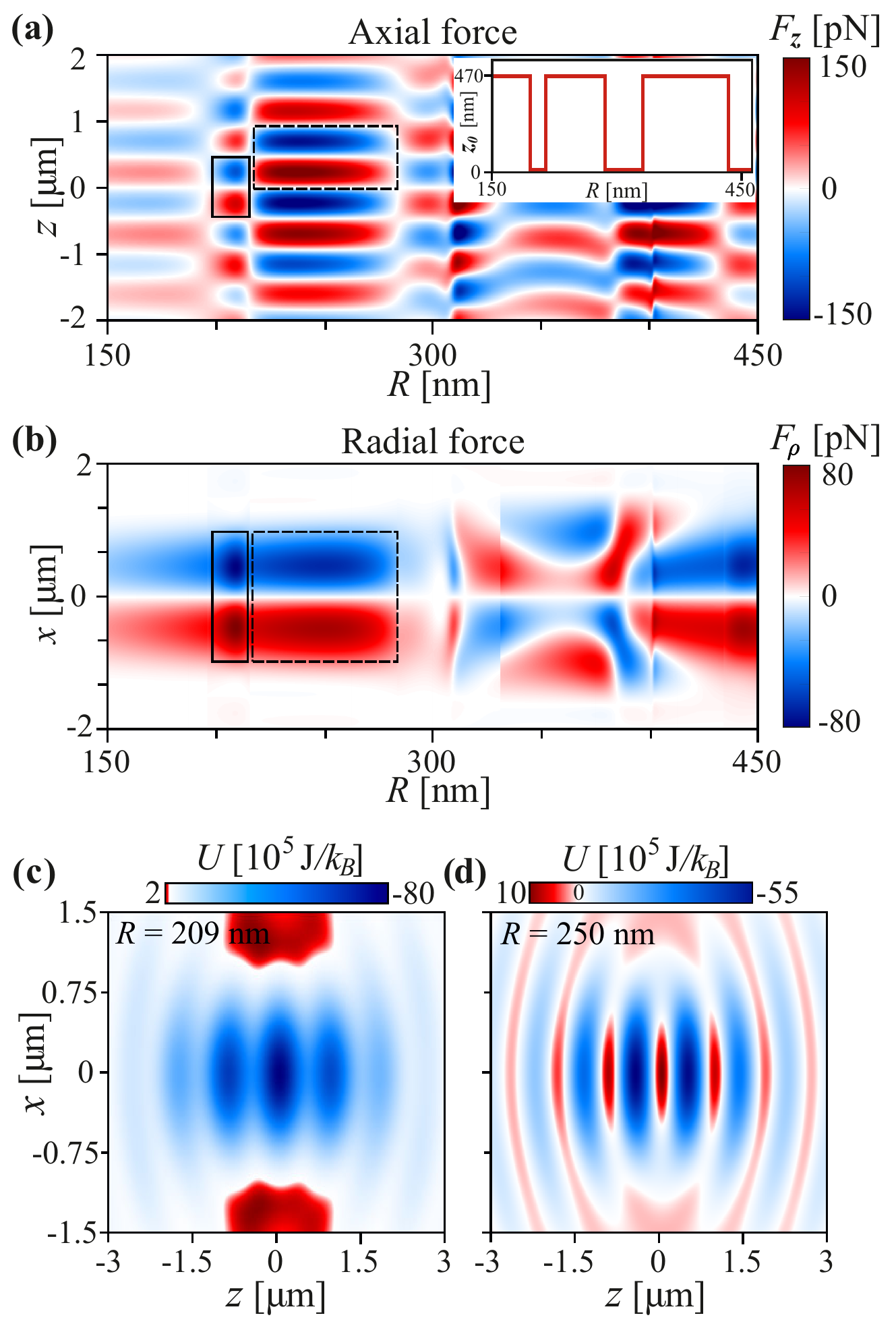}
\caption{(a) Axial force $\Fz(0,0,z)$ as a function of axial distance $z$ and radius $R$ for $\Delta\phi = \pi$ (dark focal spot). Inset shows the axial trapping position $z_0$ as a function of radius $R$. The corresponding range $R[\text{nm}] \approx (196, 215)$ and $R [\text{nm}] \approx (216, 284)$ are highlighted by solid and dashed black boxes in (a) and (b). (b) Radial force $F_\rho(x,0,z_{0})$ as a function of the radial distance $x$, radius $R$, and $z_0$ as specified in the inset of (a). (c) Optical potential $U$ for Si nanoparticles with $R = 209$~nm trapped at the dark focal point $z_0= 0$. (d) Optical potential for Si nanoparticles with $R = 250$~nm trapped at the bright spot $z_0\neq0$.
}
 \label{fig:2}
\end{figure}

To test these hypotheses, we use the optical tweezers computational toolbox (OTT)~\cite{nieminen2007optical} to calculate the total optical force on a Si nanoparticle with center-of-mass position $\rvec$ in the standing-wave configuration described above. We set the focal point of the standing wave at $z=0$ (Fig.~\ref{fig:1}). Hereafter, we make use of the paraxial approximation such that the optical force acting on a particle placed near the focus has cylindrical symmetry~\cite{novotny2012principles}. Hence, the total force is characterized by a force term along the standing-wave axis, denoted by $\Fz(\rvec)$, and a force term perpendicular to the optical axis, denoted by $\Fr(\rvec)$. In Fig.~\ref{fig:2}a we plot $\Fz(0,0,z)$ as a function of radius $R$ and axial distance $z$ for $\Delta \phi =\pi$ (dark focal spot) and using the experimental parameters given in Table \ref{tab:parameters}. Note that, $\Fz(0,0,z)$ changes its sign multiple times both as a function of $z$ due to the standing wave profile and, remarkably, as a function of radius $R$ as the trapping wavelength swipes across the different Mie resonances. 
Similar sign flips are observed for the radial force $\Fr$ as a function of radius $R$ and radial distance $x$, see Fig.~\ref{fig:2}b. For $\Delta \phi = 0$ (bright focal spot) the axial force $F_z(0,0,z)$ displays opposite attractive and repulsive regions while the radial force remains unchanged, see \cite{SM}.
Overall three-dimensional trapping can be achieved in the dark, where, in absence of the particle, light destructively interferes. As seen from the map of the optical potential $U(\rvec)$ displayed in Fig.~\ref{fig:2}c and d, the axial trapping position $z_0$ occurs either at the focal dark point $z_0=0$ for radii in the range $R[\text{nm}] \approx (196, 215)$ and in a neighbouring bright spot for radii in the range $R [\text{nm}] \approx (216, 284)$ (these radii ranges are highlighted by the black boxes in Fig.~\ref{fig:2}a and b). Hereafter, we will exclusively consider optical trapping at the focal point $z_0=0$, which requires using either $\Delta \phi = \pi$ (dark trapping) or $\Delta \phi =0$ (bright trapping) configuration depending on the particle size.

\begin{figure*}
\centering
\includegraphics[width=1\textwidth]{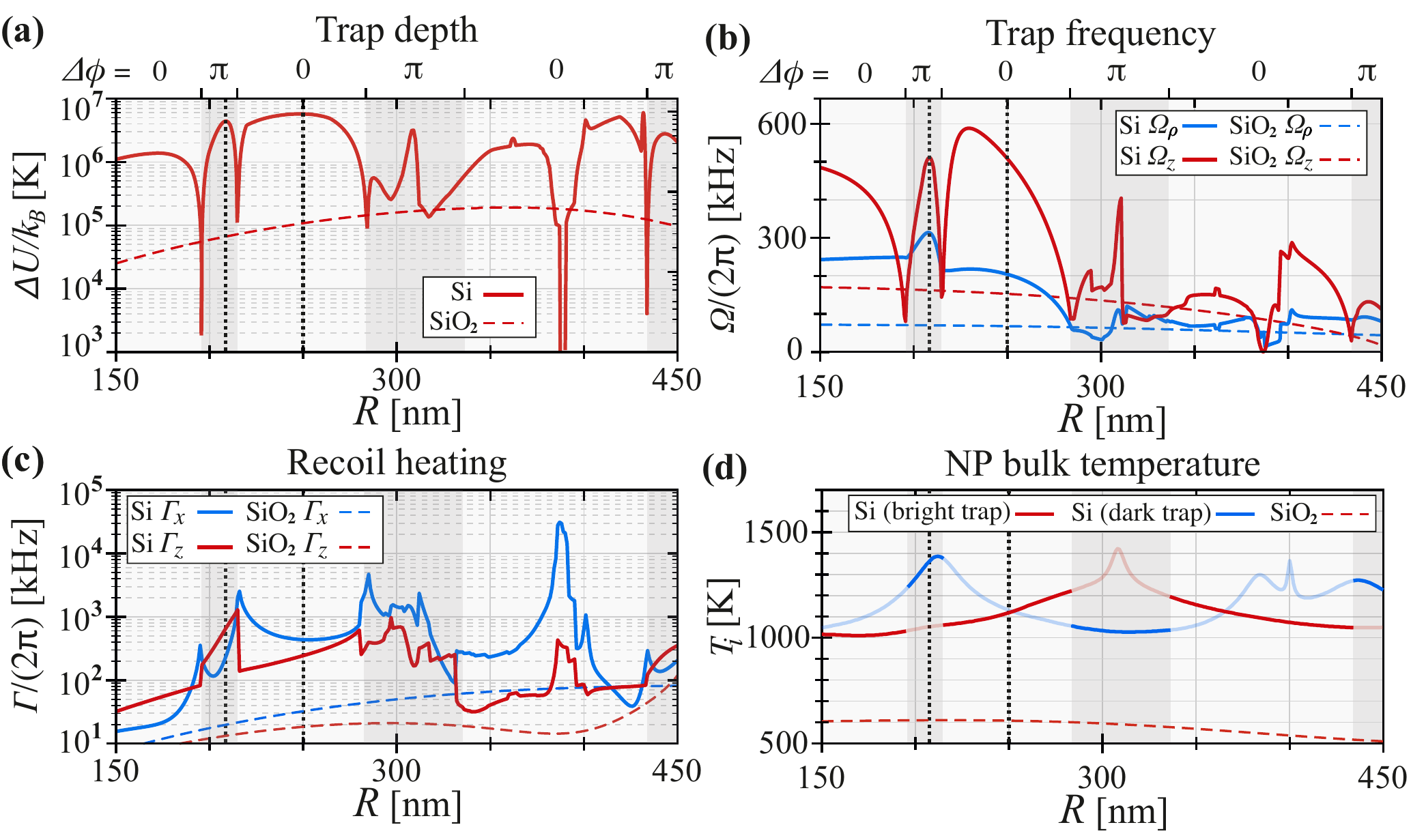}
\caption{ (a) Trap depth $\Delta U$ for $z_0 = 0$ as a function of the radius $R$ for Si (solid) and Si$\text{O}_2$ (dashed). The upper abscissa specifies which relative phase $\Delta\phi$ is used to achieve $z_0 = 0$ for Si. The black dotted lines represent the maximal trap depth at $R\approx209$~nm (dark), and $R \approx250$~nm (bright), respectively (b) Radial trap frequency $\Omega_\rho$ (blue) and axial trap frequency $\Omega_z$ (red) as a function of the radius $R$ for Si (solid) and Si$\text{O}_2$ (dashed).  (c)  Recoil heating rate $\Gamma_x$  along $x$ (blue) and recoil heating rate $\Gamma_z$ along $z$ (red) as a function of radius $R$ for Si (solid) and Si$\text{O}_2$ (dashed). (d) Internal temperature $T_i$ as a function of radius $R$ for Si in the bright trap (solid red), Si in the dark trap (solid blue), and Si$\text{O}_2$ in the bright trap (red dashed).
}
\label{fig:3}
\end{figure*}

In particle trapping, both the trap depth $\Delta U$ and the trap frequencies $\Omega_z$ and $\Omega_\rho$ are key parameters to quantify the quality of the trap. The trap depth $\Delta U$ is defined as the kinetic motional energy required for the particle to escape. The trap frequencies $\Omega_z$ and $\Omega_\rho$ are defined when the optical potential is expanded around its minimum (in our case, near the focus), namely $U(\rvec) \approx m \Omega_z^2 z^2/2 + m \Omega_\rho^2 \rho^2/2 $. In levitated optomechanics, the mechanical trap frequency along a given axis sets an important timescale for the dynamics. In the quantum regime, the trap frequency is required to be larger than the decoherence rate caused by any noise source other than laser recoil heating. Maximizing trap frequencies is thus desirable for bringing the motion of a particle into the quantum regime. In Fig.~\ref{fig:3}a we plot $\Delta U$ and in Fig.~\ref{fig:3}b $\Omega_z$ and $\Omega_\rho$ as a function of radius $R$ for a particle trapped at $z_0=0$ (the upper horizontal axis shows whether the dark $\Delta \phi=\pi$ or bright $\Delta \phi = 0$ configuration is required). For comparison, we also display with the dashed line the case of a SiO$_2$ nanoparticle for the bright configuration $\Delta \phi = 0$, which is the only possibility to trap a non-resonant dielectric particle.
In contrast to what is observed for SiO$_2$, $\Delta U$ and $\Omega_{z,\rho}$ display for Si complex $R$ dependence with regions of both enhancement and diminution. The maximum trap depth is achieved for $R=209$ nm in the dark trapping and for $R=250$ nm in the bright trapping (black dotted lines), with trap depths approximately 25 times greater than for SiO$_2$. Both axial $\Omega_z$ and radial $\Omega_\rho$ mechanical frequencies also display enhanced values, which can be more than 5 times larger than those of a SiO$_2$ nanoparticle. 
Let us remark that the complex features of Fig.~\ref{fig:3}a and b correlate to the different electric and magnetic modes supported by the particle~\cite{kivshar2017meta}, as we show in~\cite{SM}.


\begin{table}
	\begin{tabularx}{\columnwidth}{l r}
		\toprule
		\textbf{Parameters} & \text{Description }\\
		\midrule
		\midrule
		$\lambda$ = 1550 nm & laser wavelength \\
		P = 50 mW & laser power per beam \\
        \text{NA} = 0.8  & numerical aperture \\
        $n_\text{Si}(\lambda) =  3.48 + \text{i } 5.3038 \times 10^{-11}$  & refractive index of Si~\cite{degallaix2013bulk}\\
        $n_{\text{SiO}_2}(\lambda) =  1.46 + \text{i } 5 \times 10^{-9}$ & refractive index of Si$\text{O}_2$~\cite{palik1998handbook}\\
        $\rho_\text{Si} =  2330 \text{ kg m}^{-3}$ & mass density of Si\\
        $\rho_{\text{SiO}_2} =  2200 \text{ kg m}^{-3}$ & mass density of Si$\text{O}_2$\\
		\bottomrule
	\end{tabularx}
	\caption{Table of proposed experimental parameters.}
	\label{tab:parameters}
\end{table}

While so far we have focused on conservative dynamics, namely on the optical potential, the interaction of a dielectric particle with laser light also induces dissipative motional dynamics, i.e. laser light recoil heating~\cite{jain2016direct,gonzalez2019theory,maurer2022}. Recoil heating induces a linear in time increase of center-of-mass energy due to the backaction caused by the scattered light that carries information about the center-of-mass position. The recoil heating rate $\Gamma_\mu$ along the $\mu-$axis ($\mu=x,y,z$) is defined as $\partial_t E_{\mu}(t) = \Gamma_{\mu} \hbar \Omega_{\mu}$, where $E_{\mu}(t) = \langle p_\mu^2/(2m) + m \Omega_\mu^2 r_\mu^2/2 \rangle$.
The expected value represents an ensemble average over trajectories.
In the context of quantum ground-state cooling of the center-of-mass motion of a dielectric particle via optical detection, recoil heating is of paramount importance. Fig. \ref{fig:3} shows $\Gamma_{x,z}$ for Si (solid) and SiO$_2$ (dashed) particles as a function of radius $R$ calculated using recent theoretical methods~\cite{maurer2022}. We observe a complex behaviour for Si particles, while SiO$_2$ shows smooth trends expected for particles in the Rayleigh regime. In general, $\Gamma_{x,z}$ for Si exceeds SiO$_2$ for nearly all radii by up to 3 orders of magnitude. This increase is more pronounced for the axial direction. On one hand, higher $\Gamma_{x,z}$ leads to increased decoherence rates at equal power levels. On the other hand, this enhancement implies that more scattered photons carrying information about the particle position are collected, which is advantageous for active feedback cooling, as discussed below. Last but not least, we observe configurations in which $\Gamma_{x,z}$ is comparable or even larger than the trap frequencies, which is a signal of the strong optomechanical coupling regime. The possibility to enter and exploit the strong optomechanical regime in free space (i.e. without cavities) will be further investigated elsewhere.

Another critical parameter that determines the experimental feasibility of trapping in vacuum is the internal heating of the particle by laser absorption. Thus, it is important to estimate the particle's internal temperature $\Tint$ in the different optical trapping configurations under consideration. $\Tint$ in high vacuum is estimated by balancing the absorbed laser power and the power emitted by the nanoparticle, which can be calculated using Mie solutions~\cite{bohren2008absorption,landstrom2004analysis,bateman2014near}, see \cite{SM} for more details. In Fig.~\ref{fig:3}d, we show $\Tint$ for a Si nanoparticle (solid) in the bright trapping (red) and dark trapping (blue) configuration as a function of radius $R$ for the experimental numbers given in Table~\ref{tab:parameters}.
We observe three local maxima in $\Tint$ of the Si nanoparticle that align with excited Mie resonances of a different order, see \cite{SM}.  We notice that maximal values of $\Gamma_{x,z}$ coincide with increased $\Tint$. While Si nanoparticles heat up more (around a factor of 2) than SiO$_2$ nanoparticles (dashed line), the internal temperature is lower than the melting point of bulk Si ($\approx 1700$K).


\begin{figure}
\centering
\includegraphics[width=0.9\columnwidth]{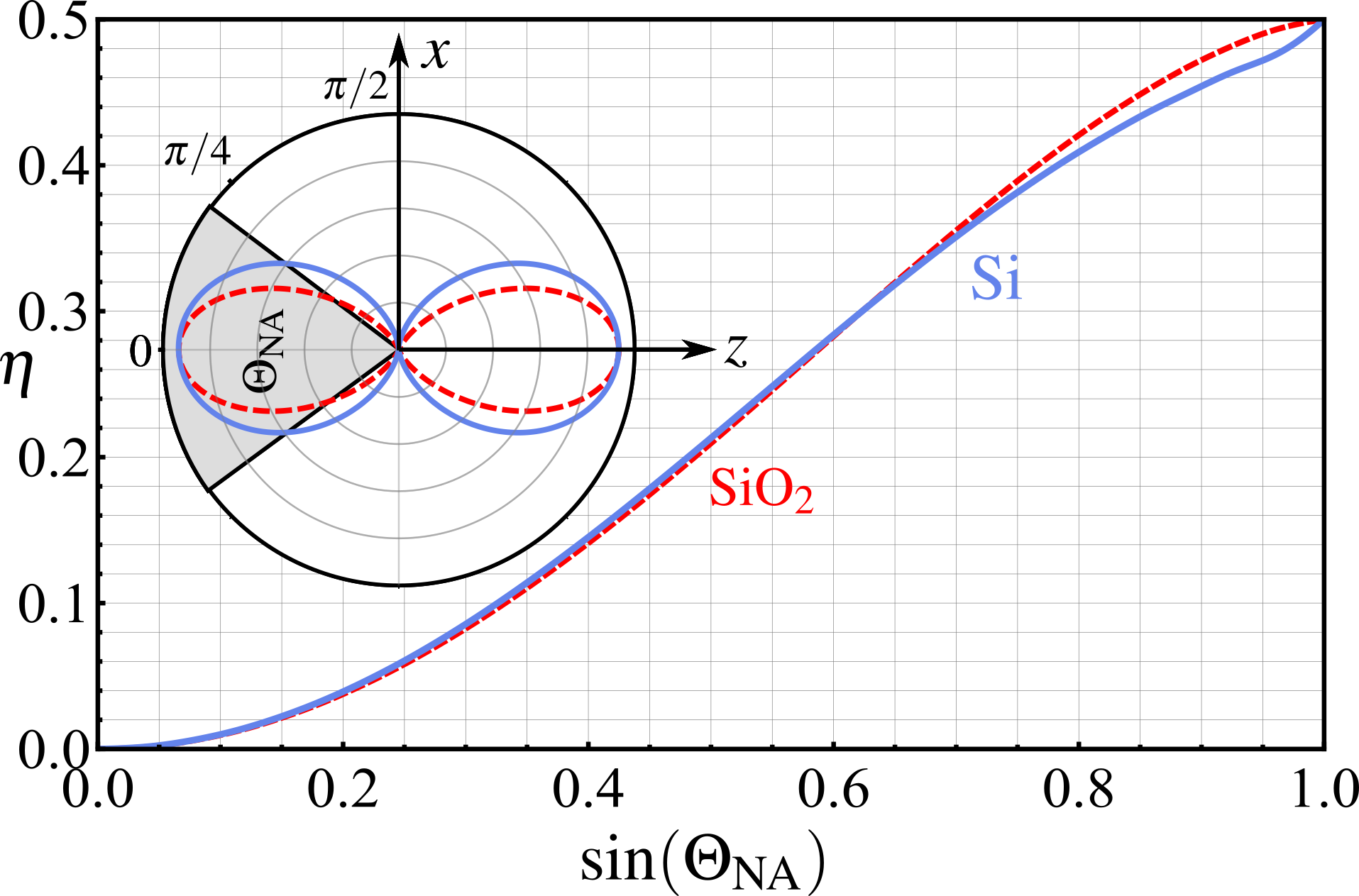}
\caption{ Detection efficiency $\eta$ along $z$ as a function of the numerical aperture NA$=\sin\Theta_\text{NA}$ for Si (solid blue) and Si$\text{O}_2$ (dashed red) at  $R=209$nm. Inset shows the (normalized) information radiation pattern along z in the x-z-plane for both Si (solid blue) and Si$\text{O}_2$ (dashed red).}
\label{fig:4}
\end{figure}


Let us now show that motional ground state cooling of a Si nanoparticle is experimentally feasible, especially in the dark trapping configuration. Assuming that laser recoil is the dominant source of motional heating, phonon occupation along a given axis, say the optical axis, $n_z$, is only governed by the detection efficiency $\eta$, with $n_z = (\sqrt{1/\eta}-1)/2<1$ \cite{tebbenjohanns2019optimal,magrini2021real,tebbenjohanns2021quantum}. $\eta$ is the ratio of detected photons that are 
scattered from the particle by either increasing or decreasing the center-of-mass kinetic energy along the $z$-axis. Hence, it is key to know the angular dependence of such scattered photons to evaluate the portion of them that can be detected and processed by the experimental configuration. This information is given by the so-called radiation patterns~\cite{tebbenjohanns2019optimal}, calculated using recent theoretical methods~\cite{maurer2022}. Fig.~\ref{fig:4} displays the radiation pattern associated with the motion along the $z$-axis in the $x$-$z$-plane for a Si nanoparticle in a dark (blue solid) and a SiO$_2$ nanoparticle in a bright trap (red dashed), and equal radii $R=209\text{ nm}$. It illustrates that, under our conditions, scattered photons feature a very similar angular pattern. The grey shaded area illustrates the collected light fraction governed by the NA. For NA$>$0.75, the reached detection efficiencies are equal for both scenarios, and for NA=0.8 the detection efficiency yields $\eta =0.41$ for Si and $\eta = 0.42$ for $\text{SiO}_2$. Hence, ground state cooling with $n = 0.28$ is in reach, as already demonstrated for bright traps in \cite{tebbenjohanns2021quantum,magrini2021real}. Let us emphasize that recoil heating for Si nanoparticles is up to 3 orders of magnitude larger than for SiO$_2$ implying that ground-state cooling can be achieved either with 3 orders of magnitude faster time scale or with up to 3 orders of magnitude less laser power. In the regimes where the recoil heating rates are comparable or larger than the trapping frequencies, other cooling methods based on the use of light pulses could be employed~\cite{bennett2016quantum}.



In summary, we have shown how Mie resonances in silicon nanoparticles introduce an additional degree of control over the dynamics of levitated mechanical oscillators. First, the higher mechanical frequencies, trap depths, and recoil heating rates as compared to standard silica particles, contribute to increased motional quantum control of nanoparticles. Second, the sign change of the particle's polarizability with the laser frequency enables trapping and center-of-mass ground-state cooling at a laser intensity minimum. We foresee that these unique properties of levitated meta-atoms will open new opportunities in levitodynamics~\cite{gonzalez2021levitodynamics}, inspired by atom optics. In particular, multi-wavelength trapping should enable the accurate control of the distance to an interface~\cite{LeKien2004atom}, critical to the study of surface forces~\cite{geraci2010short} and coupling to photonic structures \cite{forn2019ultrastrong}. Furthermore, parallel trapping of silicon meta-atoms, would allow the exploration of interactions beyond the dipole-dipole regime through higher order multipoles, such as electric or magnetic quadrupoles. 


We acknowledge valuable discussions with Yuri Kivshar, Ivan Toftul and Massimiliano Rossi. S.L. acknowledges support of Priority 2030 Federal Academic Leadership Program. This research was supported by the European Research Council (ERC) under the grant Agreement No. [951234] (Q-Xtreme ERC-2020-SyG).


%


\pagebreak
\onecolumngrid
\newpage
\begin{center} \textbf{\large Supplemental material} \end{center}
\newcommand{\beginsupplement}{
	\setcounter{table}{0}
	\setcounter{figure}{0}
	\setcounter{equation}{0}
	\setcounter{page}{1}
	\renewcommand{\thetable}{S\arabic{table}}
	\renewcommand{\thefigure}{S\arabic{figure}}
	\renewcommand{\theequation}{S\arabic{equation}}
	\renewcommand\thesection{S\arabic{section}}
}
\beginsupplement

In section~\ref{sec:S_intensity} of this supplementary material we show the intensity distributions of the
two optical configurations we consider in the manuscript, showing both intensity maxima $(\Delta\phi = 0)$ and minima $(\Delta\phi = \pi)$ at the trap origin. In section \ref{sec:S_forces} we summarize the results on the total optical force  $\textbf{F}(\rvec)$ for both optical configurations. In section \ref{sec:S_polarizability} we discuss the link between the total effective polarizability influenced by Mie resonances and the optical force. In section \ref{sec:S_scattering} we examine the scattering and absorption cross sections of the nanoparticle. Finally, in section \ref{sec:S_Tbulk} we discuss the derivation of the internal temperature of the nanoparticle. 

\section{Intensity profiles}\label{sec:S_intensity}
\begin{figure}[h]
	\centering
	\includegraphics[width=\textwidth]{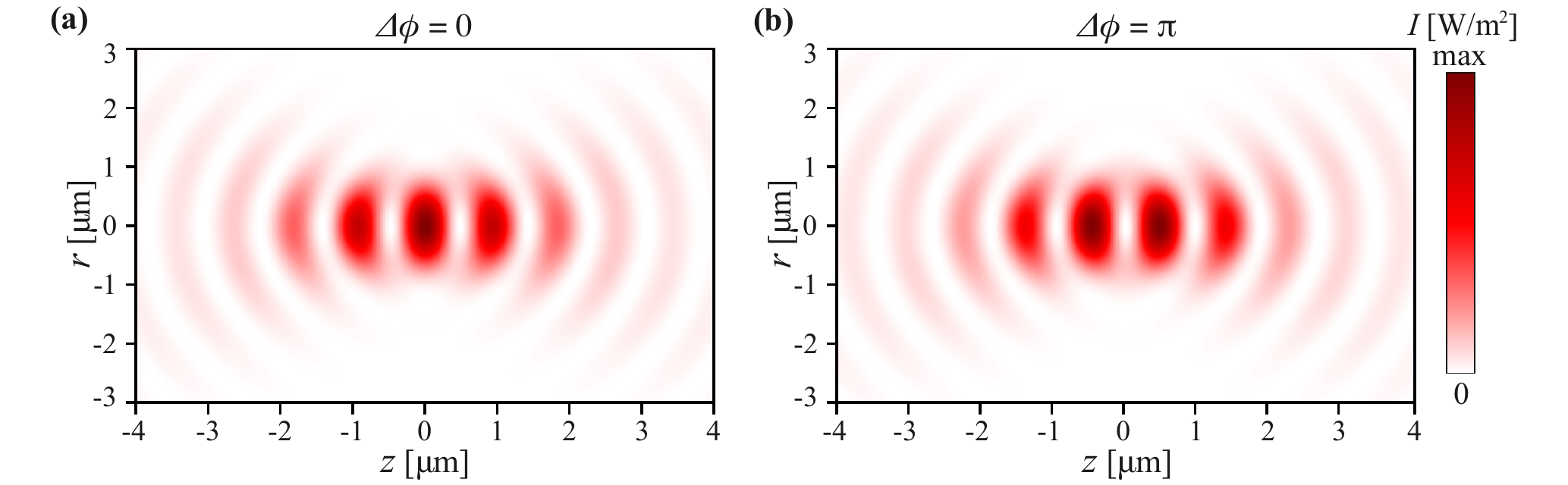}
	\caption{Intensity distribution of two focused co-polarized and counter propagating beams with (a) $\Delta \phi =0$ and intensity maximum at the origin and (b)  $\Delta \phi =\pi$ and intensity minimum at the origin.}
	\label{fig:S_intensity}
\end{figure}

We consider two coherent, counter-propagating Gaussian beams of total power $P = 100$mW, laser wavelength $\lambda=1550$nm and same transverse polarization. Each beam is focused along the optical axis $z$ by a high numerical aperture lens (NA=0.8) such that their two foci coincide and form a standing wave interference pattern with multiple intensity maxima and minima as depicted in Fig.~\ref{fig:S_intensity}. For a controlled phase difference between the two beams of $\Delta \phi =0$ ($\Delta \phi =\pi$), an intensity maximum (minimum) forms at the origin $\rvec=0$.

\section{Exerted forces}\label{sec:S_forces}
\begin{figure*}
	\centering
	\includegraphics[width=\textwidth]{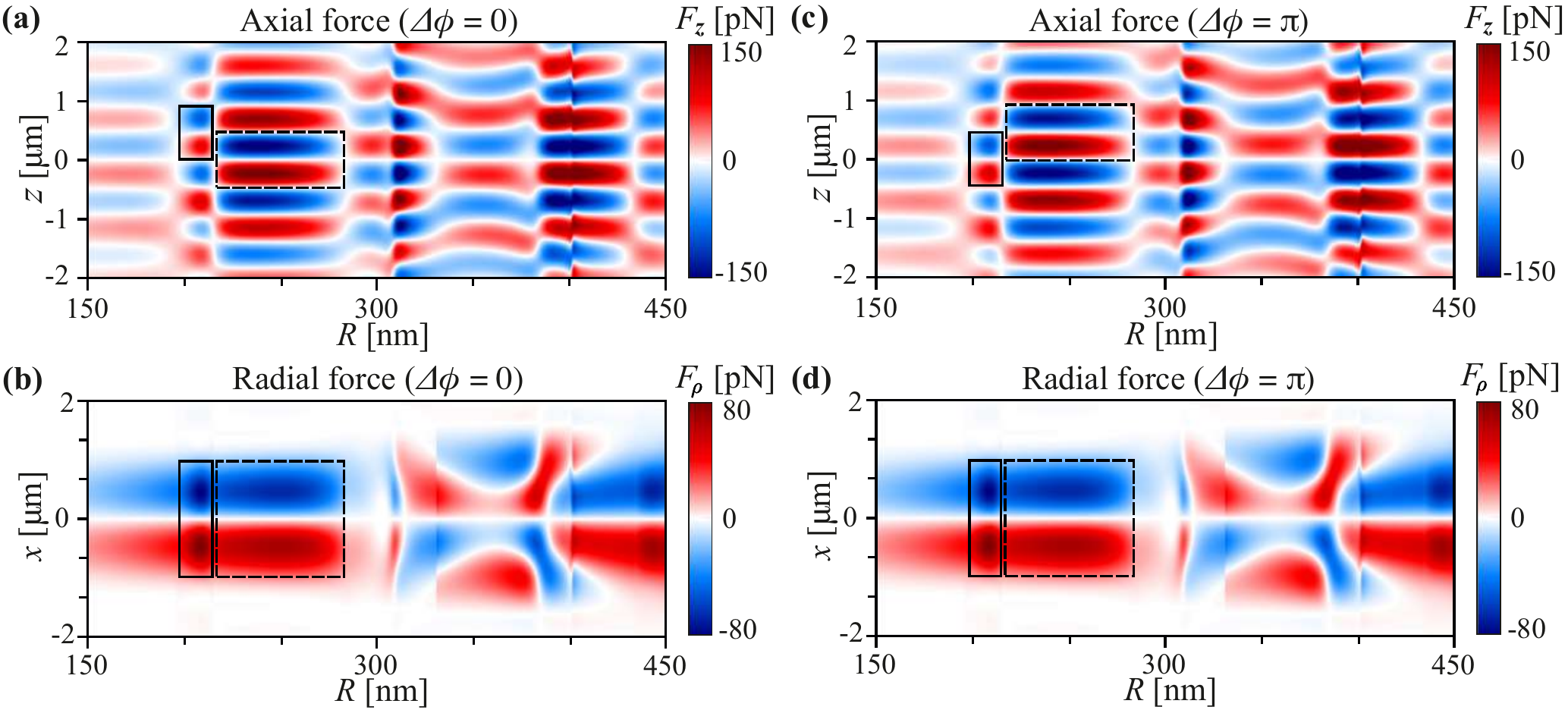}
	\caption{(a) Axial force $F_z(0,0,z)$ for $\Delta \phi=0$.  (b) Radial force $\Fr(x,0,z_0)$ for $\Delta \phi=0$. (c) Axial force $F_z(0,0,z)$ for $\Delta \phi=\pi$. (d) Radial force $\Fr(x,0,z_0)$ for $\Delta \phi=\pi$. 
		$\Delta \phi=0 $ $(\pi)$ leads to trapping in an intensity maximum (minimum) at the origin.}
	\label{fig:S_forces}
\end{figure*}

We calculate the total optical force $\textbf{F}(\rvec)$  acting on the Si nanoparticle with the optical tweezers computational toolbox (OTT) \cite{nieminen2007opticalS}. The toolbox expands the Gaussian beam into the spherical harmonic basis and calculates the electric and magnetic fields scattered by the nanoparticle using Mie theory. Taking into account the incident and scattered light fields, $\textbf{F}(\rvec)$ is calculated via the Maxwell stress tensor approach. For the complex refractive index of Si at $\lambda=1550$nm we consider $n_{\text{Si}} = 3.48+\text{i } 5.3038 \times 10^{-11}$~\cite{degallaix2013bulkS} and interpolate 
$n_{\text{SiO}_2} = 1.46 + \text{i }5 \times 10^{-9}$ for SiO$_2$ ~\cite{palik1998handbookS}.

Hereby, we consider two different trapping scenarios: First, an intensity maximum at the origin for $\Delta \phi =0$ (see Fig.~\ref{fig:S_intensity}a) giving rise to the optical forces $\textbf{F}(\rvec)$ displayed in Fig.~\ref{fig:S_forces}a and b, and an intensity minimum at the origin for $\Delta \phi =\pi$ (see Fig.~\ref{fig:S_intensity}b) with optical forces $\textbf{F}(\rvec)$ as depicted in Fig.~\ref{fig:S_forces}c and d.
The axial force $F_z(0,0,z)$ displays in both cases frequent sign changes with $R$ and $z$. For the two different $\Delta \phi$ we find forces of opposite sign, as can be seen in Fig.~\ref{fig:S_forces}a and c. This enables a dark trap center at the origin for $\Delta \phi =\pi$.
The radial force $\Fr(x,0,z_0)$ shows a similar frequent sign change with $R$, but the force direction is the same for both $\Delta \phi$. Due to the absence of interference in the radial direction, we observe at most one trap centre at $\rvec=0$; except for radii in the range $R \approx [330\text{nm},390\text{nm}]$, where MR give rise to additional trap centers at $\rvec\neq 0$. We attribute this effect to the interplay of several mulitpole MR at these particle sizes (see Fig.~\ref{fig:S_polarisability}a). We extract $\Delta U$ as shown in Fig.~3a in the main text from the potential energy  $U(\rvec)= -\int_{-\infty}^{\rvec} \textbf{F}(\rvec')d\rvec'$, where $\rvec = [x,y,z]$ is the coordinate of the point where the potential is evaluated.

\section{Influence of the effective polarizability on trap characteristics}\label{sec:S_polarizability}

\begin{figure*}
	\centering
	\includegraphics[width=1\textwidth]{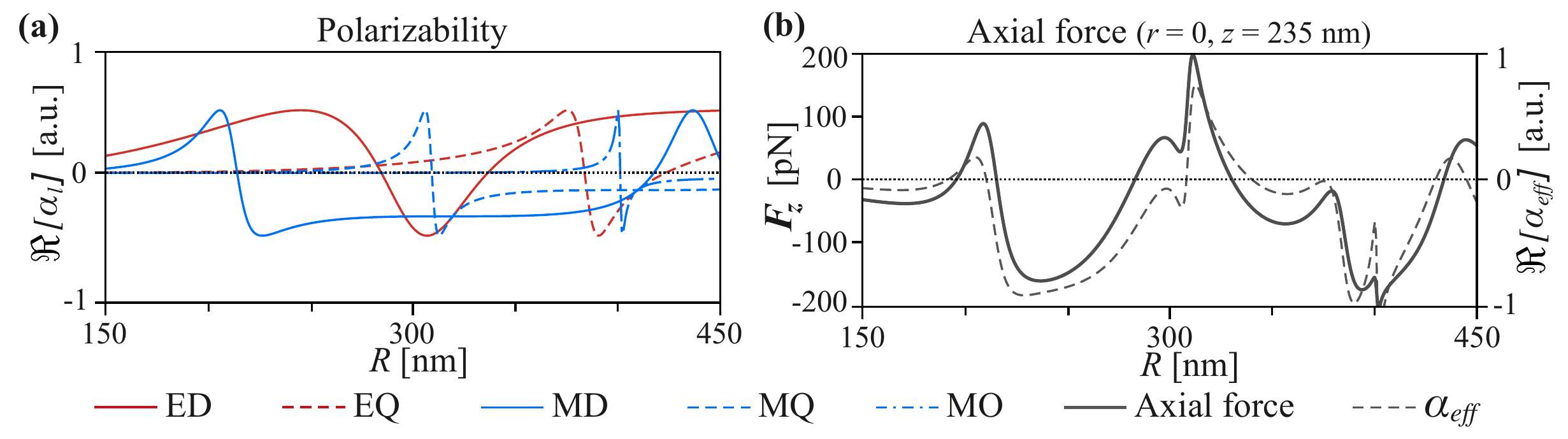}
	\caption{\label{fig:4} Effective polarizability and optical forces (a) Real part of the polarizabilities  $\Re{[\alpha_{l}]}$ associated with the ED (red solid), EQ (red dashed), MD (blue solid), MQ (blue dashed) and MO (blue dash-dotted) moments depending on the Si nanoparticle radius. (b) $F_z(0,0,235\text{nm})$ (black solid) and $\alpha_{\text{eff}}$ (black dashed) as a function of the Si nanoparticle radius $R$.}
	\label{fig:S_polarisability}
\end{figure*}

As pointed out in the manuscript, we attribute the modifications of $\textbf{F}(\rvec)$, $\Delta U$, $\Omega_{\rho,z}$, $\Gamma_{x,z}$ and $\Tint$ at certain $R$ to the excitation of optical MR. We calculate polarizabilities $\alpha_l$ associated with excitation of the MR of different order $l$ (corresponding to different electric and magnetic multipolar contributions)~\cite{chen2011opticalS, evlyukhin2016opticalS}. The real part of those multipole polarizabilities are depicted in Fig.~\ref{fig:S_polarisability}a. Here, MD (ED) is the magnetic (electric) dipole,  MQ (EQ) is the magnetic (electric) quadrupole, and MO is the magnetic octupole. Note that the real part of the multipole polarizabilities is proportional to the imaginary part of the Mie scattering coefficients~\cite{evlyukhin2016opticalS}. 
It is well-known that in the simplified dipole approximation, $\textbf{F}(\rvec)$ is proportional to $\Re{[\alpha_{\text{ED}}]}$. By cancelling the scattering force in counter-propagating beams, we suggest that the gradient force of $\textbf{F}(\rvec)$ is proportional to the real part of an effective polarizability $\alpha_{\text{eff}}$ which can be presented as a linear combination of  multipole polarizabilities $\alpha_{l}$. To support our hypothesis, we approximate $\Re{[\alpha_\text{eff}]}$ by a linear combination of $\Re{[\alpha_{l}]}$ \cite{chen2011opticalS} such that it fits the axial force $F_z$, namely
\begin{eqnarray}
	\Re{[\alpha_\text{eff}]} = (\Re{[\alpha_{\text{MD}}]} - \Re{[\alpha_{\text{ED}}]}) - (\Re{[\alpha_{\text{MQ}}]} - \Re{[\alpha_{\text{EQ}}]}) + \Re{[\alpha_{\text{MO}}]}.
	\label{eq:alpha_eff}
\end{eqnarray}
In Fig.~\ref{fig:S_polarisability}b, we plot $F_z$ for $\Delta\phi = 0$ as a function of the radius $R$ at the specific point $\rvec= (0,0,235\text{nm})$, where $F_z$ reaches the maximum value (see Fig.~\ref{fig:S_forces}a). Comparing $F_z$ (black solid) with $\Re{[\alpha_\text{eff}]}$ (black dashed) in Fig.~\ref{fig:S_polarisability}b, we find qualitative agreement in their behaviour with $R$, enabling us to link qualitatively certain MR to optical force modifications. The discrepancies between $F_z$ (Fig.~\ref{fig:S_polarisability}b, black solid) and $\Re{[\alpha_\text{eff}]}$ (Fig.~\ref{fig:S_polarisability}b, black dashed) are caused by the fact that the scattered electric field expanded into a series of $\alpha_{l}$ contributes quadratically to $\textbf{F}(\rvec)$~\cite{farsund1996forceS}, and the linear combination gives a limited approximation. 
For $150\text{nm}\leq R\leq 450\text{nm}$, the dominant terms are the ED (solid red curve), MD (solid blue curve), EQ (dashed red curve), MQ (dashed blue curve) and MO (dash dotted blue curve). 

\section{Scattering and absorption cross section}\label{sec:S_scattering}

\begin{figure*}[h]
	\centering
	\includegraphics[width=1\textwidth]{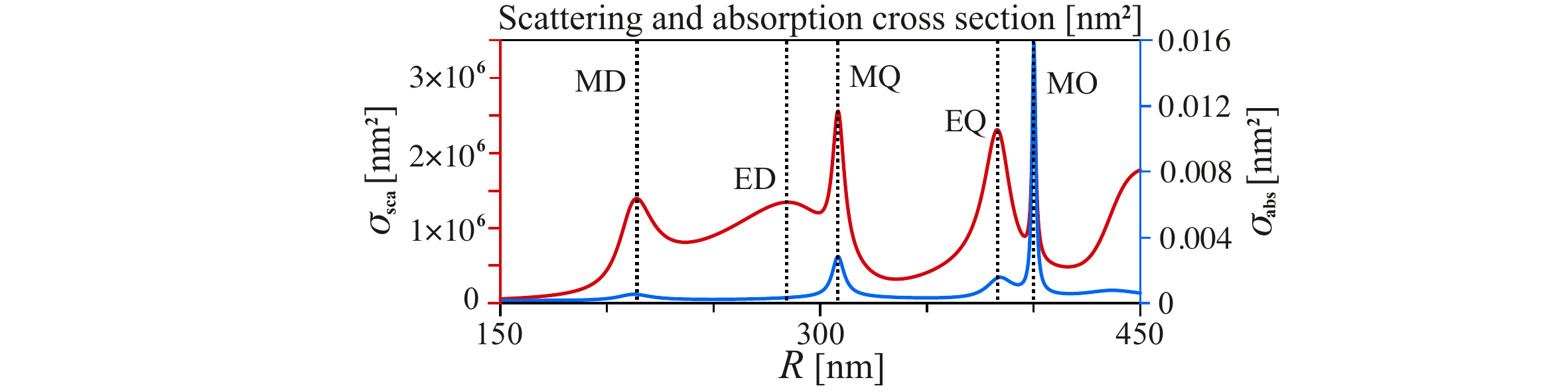}
	\caption{Scattering (red) and absorption (blue) cross sections of Si nanoparticles as a function of $R$ at $\lambda=1550$nm. The dashed lines represent the resonances at $R_{\text{MD}}=214$~nm, $R_{\text{ED}}=285~$nm, $R_{\text{MQ}}=308$~nm, $R_{\text{EQ}}=383$~nm, and $R_{\text{MO}}=400$~nm. 
	}
	\label{fig:S_scattering}
\end{figure*}

In Fig.~\ref{fig:S_scattering} we show the scattering cross section $\sigma_\text{sca}$ and absorption cross section $\sigma_\text{abs}$ for a single plane wave propagating along the $z-$axis as a function of the nanoparticle radius $R$. We observe that the cross section shares many features with $\Re{[\alpha_\text{eff}]}$ in Fig.~\ref{fig:S_polarisability}. The dashed lines highlight the individual MR. In particular, we recognize resonances at $R_{\text{MD}}=214$~nm, $R_{\text{ED}}=285~$nm, $R_{\text{MQ}}=308$~nm, $R_{\text{EQ}}=383$~nm and $R_{\text{MO}}=400$~nm, where MD (ED) is the magnetic (electric) dipole,  MQ (EQ) the magnetic (electric) quadrupole and MO the magnetic octupole. The absorption cross section $\sigma_\text{abs}$ is calculated using Mie theory as the difference $\sigma_\text{abs}=\sigma_\text{ext}-\sigma_\text{sca}$ between extinction cross section $\sigma_\text{ext}$ and scattering cross section $\sigma_\text{sca}$~\cite{bohren2008absorptionS}, where
\begin{eqnarray}
	\sigma_\text{sca}&=&\frac{2\pi}{k^2}\sum^{\infty}_{l=1}(2l+1)(|a_l|^2+|b_l|^2), \\
	\sigma_\text{ext}&=&\frac{2\pi}{k^2}\sum^{\infty}_{l=1}(2l+1)\Re(a_l+b_l),
	\label{eq:CS}
\end{eqnarray}
and where $k=2\pi/\lambda$ is the wavenumber, and $l$ is the order of the electric $a_l$ and magnetic $b_l$ Mie coefficients \cite{bohren2008absorptionS}.

\section{Nanoparticle internal temperature calculation}\label{sec:S_Tbulk}

We determine the steady-state internal temperature $T_{i}$ of a nanoparticle in the laser field by minimizing the difference of absorbed laser power and the power emitted by the nanoparticle, that is
\begin{equation}
	T_i = \operatorname*{argmin}_{T_i} \left\lbrace  P_\text{abs} + \int  \left[ P_{BB}(T_\text{env},\omega)- P_{BB}(T_{i},\omega)\right]d\omega \right\rbrace,
	\label{eq:power}
\end{equation}
where $P_\text{abs}$ is the absorbed laser power, $ P_{BB}(T_\text{env},\omega)$ is the power spectral density of the absorbed environmental black body radiation at room temperature $T_\text{env}$ responsible for the nanoparticle heating and $P_{BB}(T_i,\omega)$ is the power spectral density of the nanoparticle emitted black body radiation responsible for cooling. The absorbed laser power is calculated by means of the optical tweezer toolbox (OTT) as the difference $P_\text{abs}=P_\text{out}-P_\text{in}$ between the power $P_\text{out}$ leaving the system and the incident laser power $P_\text{inc}$. The incident power is equal to the power of the incoming co-polarized beams and expressed only through the incident electromagnetic field as $P_\text{inc} = \frac{1}{2}\int_S(\mathbf{E}_\text{inc}\times\mathbf{H}_\text{inc}^*)\cdot d\mathbf{S} = 100$~mW. Here, the integration is taken on the plane perpendicular to the propagation directions. The outgoing power is obtained from the total field, i.e. a superposition of incident and scattered fields, and reads $P_\text{out} = \frac{1}{2}\oint_S[(\mathbf{E}_\text{inc}+\mathbf{E}_\text{scat})]\times(\mathbf{H}_\text{inc}^*+\mathbf{H}_\text{scat}^*)\cdot d\mathbf{S}$. The integration is performed on the closed surface surrounding the nanoparticle. The nanoparticle is placed at the origin ($\rvec=0$), and the phase difference between the counter-propagating of the beams takes on the values of $\Delta \phi = 0$ and $\pi$ giving rise to the bright and dark trapping (Fig.~\ref{fig:S_intensity}a and b) respectively. The power spectral density of the black body radiation reads
\begin{eqnarray}
	P_{BB}(T,\omega) = \frac{\hbar\omega^3}{\pi^2c^2[\exp(\hbar\omega/k_B T)-1]}\sigma_\text{abs}(\omega),
	\label{eq:BBradiation}
\end{eqnarray}
where $c$ denotes the speed of light in vacuum and $k_B$ denotes the Boltzmann constant. The absorption cross section $\sigma_\text{abs}(\omega)$ depends on the complex refractive index of the nanoparticle. Thus $P_{BB}(T,\omega)$ becomes refractive index-dependent across the entire spectrum of the black body radiation. For the Si nanoparticle, we merge the refractive index spectra from Refs.~\cite{degallaix2013bulkS,schinke2015uncertaintyS,chandler2005highS}, and for the SiO$_2$ nanoparticle, we use the interpolated data from Ref.~\cite{palik1998handbookS}.

\section{Trap characterization for silicon and silica nanoparticles at identical recoil heating rates }\label{sec:comparison}

In the following, we compare the trap performance when the recoil heating rates $\Gamma_{\textrm{Si}}$ and $\Gamma_{\textrm{SiO}_{2}}$  experienced by Si and SiO$_2$ nanoparticles are equal. We discussed $\Gamma_{\textrm{Si}}$ in Fig. 3(c) of the main manuscript assuming $P_{\textrm{las}}=0.05$~W.
Since $\Gamma \propto   P_{\textrm{las}} / \Omega$ and $\Omega \propto \sqrt{P_{\textrm{las}}}$ ~\cite{gonzalez2019theoryS}, we find overall that $\Gamma\propto\sqrt{P_{\textrm{las}}}$. This expression allows us to extract $P_{\textrm{las}}$ required to balance the recoil heating rates as $ P_{\textrm{las}}^{\textrm{SiO}_{2}} = P_{\textrm{las}}^{\textrm{Si}} (\Gamma_{\textrm{Si}}/\Gamma_{\textrm{SiO}_2})^2$.

Let us consider the dark (bright) trapping performance with $R = 209$~nm ($R = 250$~nm) in the $x$ and $z$ direction. 
For $R = 209$~nm, the respective recoil heating rates are equal at  $P_{\textrm{las}}^{\textrm{SiO}_{2}, x} = 16.6$~W and $P_{\textrm{las}}^{\textrm{SiO}_{2}, z} = 56.9$~W, whereas for $R = 250$~nm this happens at $P_{\textrm{las}}^{\textrm{SiO}_{2}, x} = 26.9$~W and $P_{\textrm{las}}^{\textrm{SiO}_{2}, z} = 2.7$~W. For all cases, $P_{\textrm{las}}^{\textrm{SiO}_{2}}$ exceeds $P_{\textrm{las}}^{\textrm{Si}}= 0.05$~W by orders of magnitude. For the following analysis, we select $P_{\textrm{las}}^{\textrm{SiO}_{2}} = 2.7$~W.

\begin{figure*}[h]
	\centering\includegraphics[width=1\linewidth]{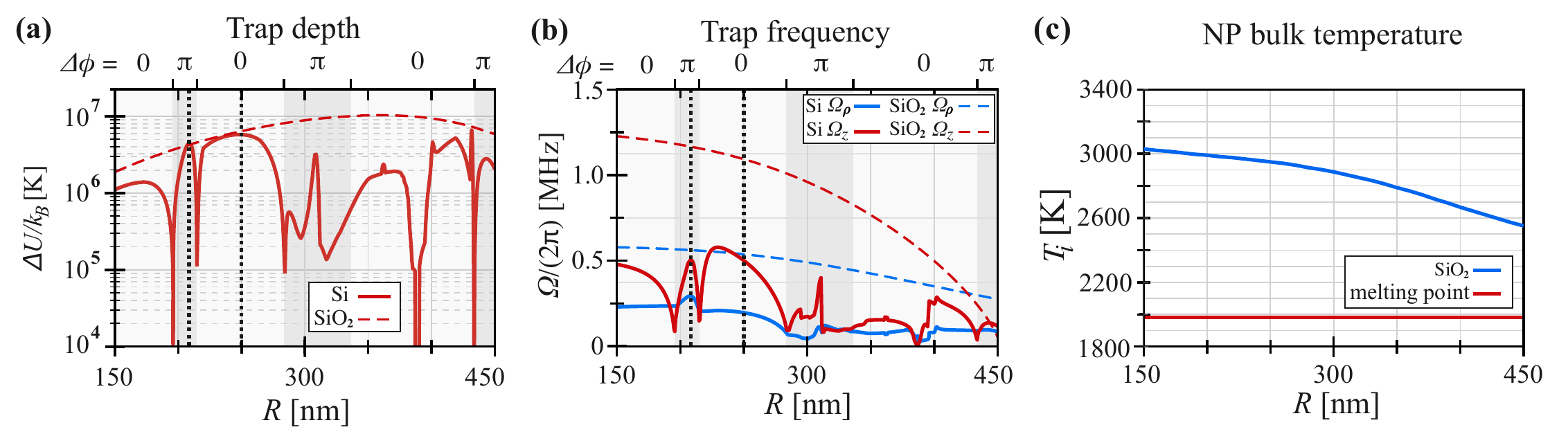} 
	\caption{(a)~Trap depth of silicon (solid) and silica (dashed) nanoparticles as a function of $R$. (b)~Radial (blue) and axial (red) trap frequencies as a function of $R$ for silicon (solid) and silica (dashed) nanoparticles. The black dotted lines indicate the local maxima of the trap depth of silicon nanoparticle in the bright ($R\approx 209$~nm) and dark ($R\approx 209$~nm) traps. (c)~Internal temperature of the silica nanoparticle as a function of $R$. The red line represents the melting temperature of bulk silica ($\approx1986$~K). \label{figR1}}
	\label{fig:5}
\end{figure*}

Figure~\ref{figR1} shows the trap depth $\Delta U$, trap frequency $\Omega$ and internal temperature $T_i$ as a function of $R$. $P_{\textrm{las}}$ for each counter-propagating beam is fixed at 0.05~W and 2.7~W for Si and SiO$_2$, respectively. Other parameters of the system remain the same as in the manuscript. Clearly, increasing $P_{\textrm{las}}$ raises both $\Delta U$ and $\Omega$ of SiO$_2$ to values significantly higher than those of Si (Figure~\ref{figR1}(a,b)). However, $T_i$ of the SiO$_2$ nanoparticle increases much beyond the melting point for such a high laser power. Note that, our internal temperature calculations consider only the linear optical losses of silica,  disregarding the losses caused by optical non-linearities that arise at such high laser power levels. 

%

\end{document}